\documentclass[a4paper]{jpconf}
\pdfoutput=1
\usepackage{graphicx}
\usepackage{amsmath}
\usepackage{hyperref}
\usepackage{subcaption}
\begin{document}

\title{Implementation of Allan Standard Deviation Technique in Stability Analysis of 4C31.61 Quasar Position}

\author{Jessica Syafaq Muthmaina$^{[1]}$, Ibnu Nurul Huda$^{[2]}$, Dwi Satya Palupi$^{[1]}$}

\address{$^{[1]}$Department of Physics, Faculty of Mathematics and Natural Sciences, Universitas Gadjah Mada, Sekip Utara, BLS 21, Yogyakarta 55281, Indonesia}
\address{$^{[2]}$Research Center for Computation, National Research and Innovation Agency (BRIN), Jalan Raya Jakarta-Bogor, 16915 Bogor, Indonesia}

\ead{syafaqmuth@mail.ugm.ac.id, ibnu.nurul.huda@brin.go.id, dwi\_sp@ugm.ac.id}

\begin{abstract}
The International Celestial Reference Frame (ICRF) plays an important role in astronomy and geodesy. The realization of ICRF is based on the position of thousands of quasars observed using the Very-Long Baseline Interferometry (VLBI) technique. Better quality of ICRF is achieved when the position of the quasars is stable. In this study, we aim to analyze the stability of one of the quasars in ICRF called 4C31.61 (2201+315). We performed VLBI data analysis by using Vienna VLBI and Satellite Software (VieVS) to get the position of the quasar. We also used the data of the quasar's position from the Paris Observatory Geodetic VLBI Center. We examined the stability of the quasar position by using the Allan standard deviation technique. We found that the quasar 4C31.61 (2201+315) has a stable position with the dominance of white noise across the majority of time scales.
\end{abstract}

\section{Introduction}

Very Long Baseline Interferometry (VLBI) is an observational technique that use some widely separated radio telescopes to observe astronomical radio source simultaneously. VLBI has been measuring radio source positions since in the mid-1960s \cite{sovers1998astrometry}. VLBI has contributed to several domains. For instance, it is the only technique that can monitor the Earth Orientation Parameters completely. VLBI has contributed also to the realization of Celestial and Terrestrial Reference Frame \cite{schuh2012vlbi}. Indonesia has several ongoing radio telescope construction projects, such as Bosscha Observatory, located in Bandung city, and National Observatory, located in Timau Mountain. These telescopes are targeted to join the global Very Long Baseline Interferometry (VLBI) network \cite{huda2021measuring}.

A celestial reference frame has been used to observe the position of celestial objects with high precision. Originally, the International Celestial Reference Frame (ICRF-1) was proposed in 1998 to replace FK5 \cite{ma98}. This ICRF is based on Very Long Baseline Interferometry (VLBI) observation over 212 extragalactic radio sources distributed over the entire sky. ICRF-1 was refined by ICRF-2 through incorporating more extragalactic radio sources \cite{ICRF2}. ICRF-3 was released in 2020 by offering increased precision and encompassing a greater number of extragalactic radio sources compared to its predecessors \cite{ICRF3}. 

One of the important issues in the realization of ICRF is the position stability \cite{ICRF3}. It has an implication for the quality of VLBI products. Previous studies have dealt with this problem \cite{gontier2001stability,feissel2003selecting,gattano2018extragalactic}. In this study, we focused on the use of recent VLBI observation for studying the quasar position stability. We analyzed the stability of one of the quasars named 4C31.61 (2201+315) by using the Allan standard deviation technique.

\section{Data and Method}

The position of the quasar is given in Right Ascension ($\alpha$) and Declination ($\delta$). Here we considered the quasar position from two sources. First, we adopted the data from the Paris Observatory VLBI Center (\url{https://syrte.obspm.fr/~lambert/ivsopar/}). Second, we conducted a VLBI analysis by using software Vienna VLBI Software or VieVS \cite{bohm2018vienna}. The VLBI analysis was done by estimating the source positions as session-wise parameters. The reduced least squares adjustment technique was used to estimated the parameters. We used 6342 VLBI sessions in our calculation. We used ICRF-3 \cite{ICRF3} and ITRF-2020 \cite{altamimi2023itrf2020} as our Celestial and Terrestrial Reference Frame respectively.  TPXO72 \cite{egbert2002efficient} was considered for the tidal oceanic loading model and APL-VIENNA \cite{wijaya2013atmospheric} for the atmospheric tidal and non-tidal loading model. The Earth Orientation Parameters were estimated with respect to the EOP C04 series \cite{bizouard2019iers}. Here we used 33 years of VLBI observation data ranging from 1990 to 2023.

We analyzed the stability of quasar 4C31.61 by using the Allan standard deviation technique \cite{allan66}. It is a technique that detect the type of the noise in the time series as a function of the time scale. The overlapping Allan standard deviation is given as follows \cite{gattano2018extragalactic}:
\begin{equation}
    \sigma^2(\tau) = \frac{1}{M-2m+1} \sum^{M-2m+1}_{j=1}\frac{1}{2}\left(\frac{1}{m}\sum^{j+m-1}_{i=j}(\Bar{y}_{i+m}-\Bar{y}_i)\right)^2
\end{equation}
where $M$ is the number of possible windows in the time series, $\tau = m\tau_0$ is the time scale, $\tau_0$ means the mean sampling, $m$ is the integer, $\Bar{y}$ mean of sample with considered time interval $\tau$. The module Allantools (\url{https://github.com/aewallin/allantools}) was used to calculate the overlapping Allan deviation.

The stability of quasar positions was analyzed by identifying the dominant type of noise in Allan variance plots. Different types of noise were categorized based on the slope of each plot (in logarithmic scale). There are three noise types as follows: white noise or stable noise with a slope lower than -0.25, flicker noise with a slope between -0.25 and 0.25, and random walk or unstable noise with a slope greater than 0.25.

\section{Results}

\begin{figure}
    \centering
    \includegraphics[width=0.85\textwidth]{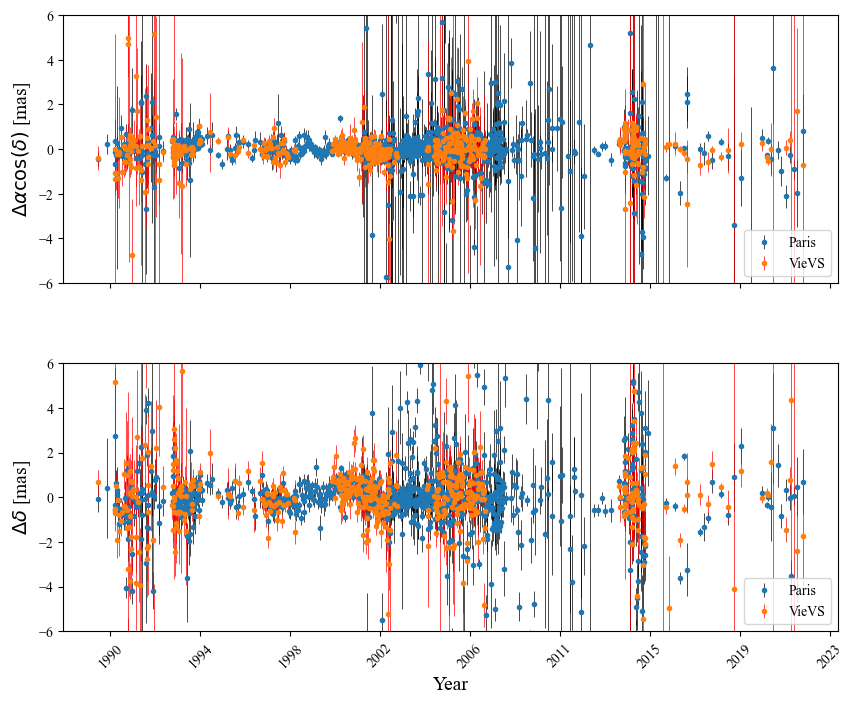}
    \caption{Comparison of 4C31.61 relative position given by our VLBI analysis (VieVS) and Paris Observatory Geodetic VLBI Center (Paris). The position is depicted relative to the reference value.}
    \label{fig:Comparison}
\end{figure}

\begin{figure}
    \centering
    \begin{subfigure}[b]{0.9\textwidth}
    \centering
    \includegraphics[width=1\hsize]{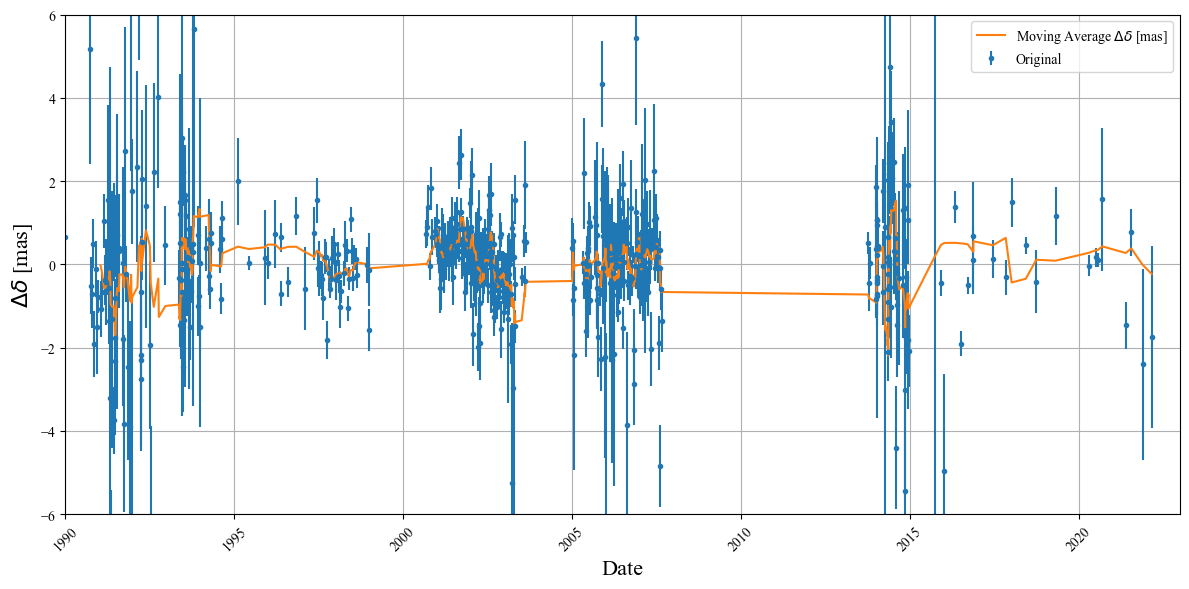}
    \caption{}
         \label{fig:VieVS_dec}
     \end{subfigure}
     \begin{subfigure}[b]{0.9\textwidth}
    \centering
    \includegraphics[width=1\hsize]{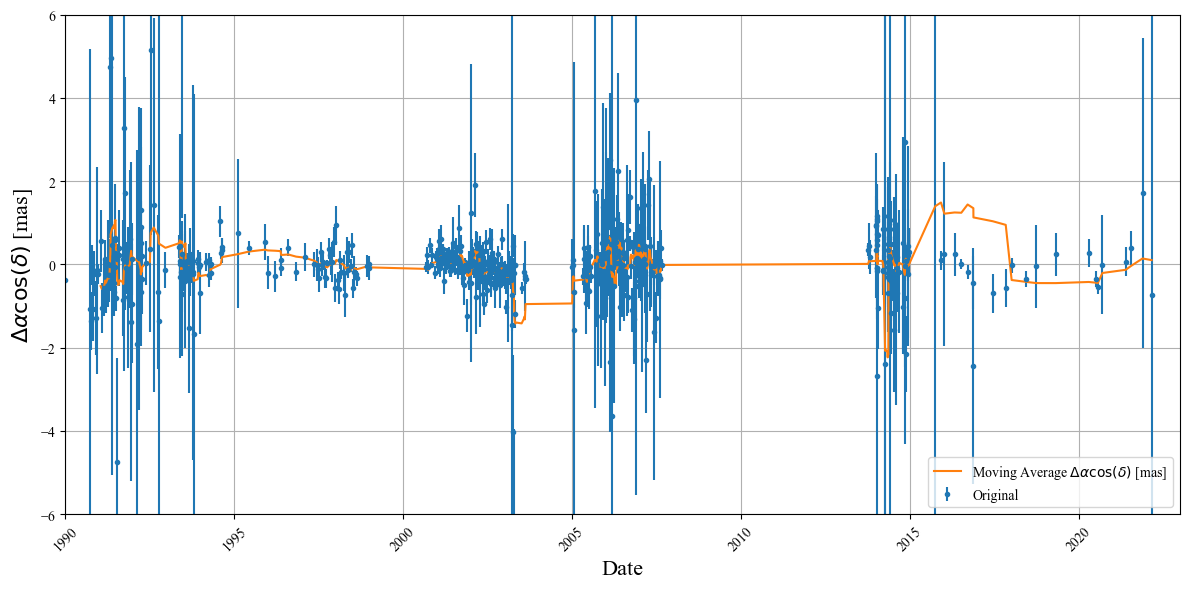}
    \caption{}
         \label{fig:VieVS_ra_cos_dec}
     \end{subfigure}
\caption{\label{fig:Mov_av} Comparison of original data with the data after conducting moving average technique. The position is depicted relative to the reference value.}
\end{figure}

Figure \ref{fig:Comparison} shows the comparison of the 4C31.61 position derived from our VLBI analysis and Paris Observatory Geodetic VLBI Center. The position was plotted with respect to the reference value where $\alpha = 330.8123990589822^o$ and $\delta = 31.7606305185240^o$. Generally, it shows that our results have a good agreement with Paris. However, we noted that our VLBI analysis produced a gap in some periods, e.g. 2007 - 2014, which caused by our technical problem during data processing. Next, we processed the time series by removing the outliers and applying the moving average. Figure \ref{fig:Mov_av} shows the results for the data from our VLBI analysis. We also conducted a similar procedure for Paris data.

\begin{figure}
    \begin{subfigure}[b]{0.49\textwidth}
    \centering
    \includegraphics[width=1\hsize]{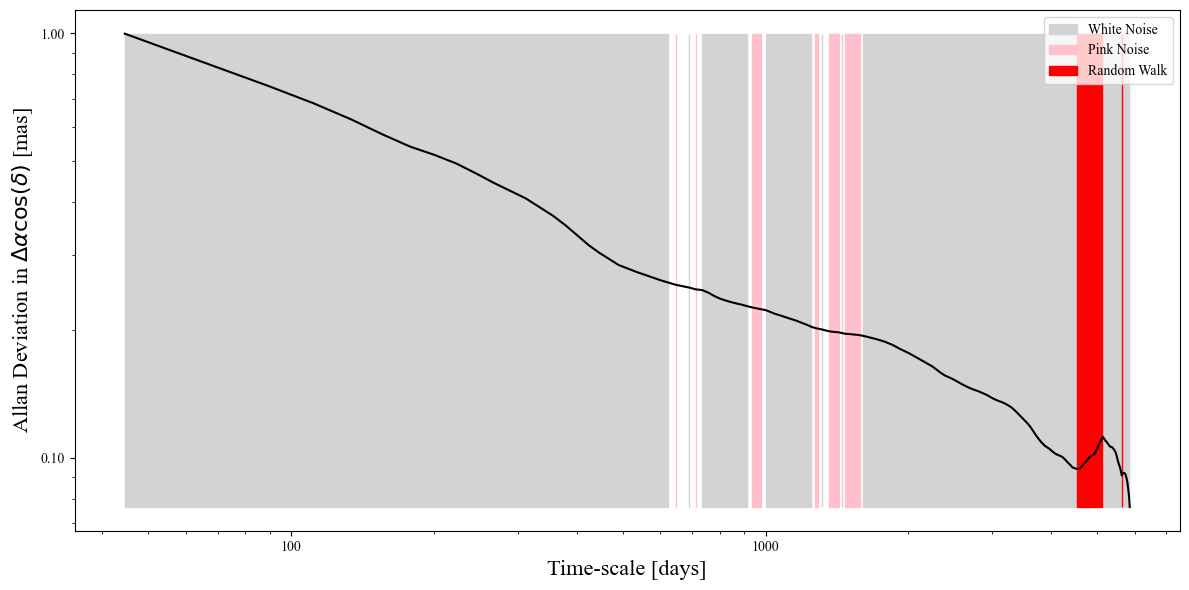}
    \caption{}
         \label{fig:VieVS_Allan_ra_cos_dec}
     \end{subfigure}
     \begin{subfigure}[b]{0.49\textwidth}
    \centering
    \includegraphics[width=1\hsize]{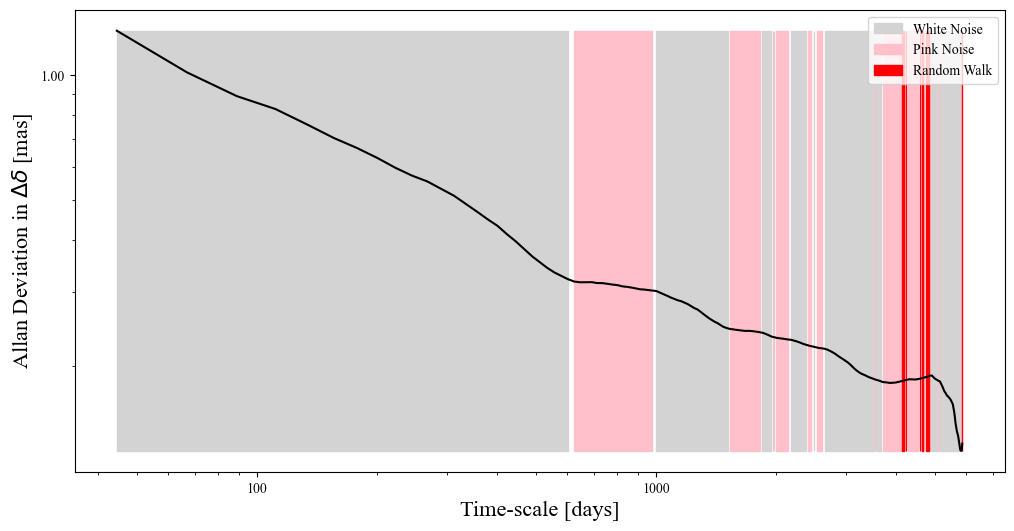}
    \caption{}
         \label{fig:VieVS_Allan_dec}
     \end{subfigure}
      \begin{subfigure}[b]{0.49\textwidth}
    \centering
    \includegraphics[width=1\hsize]{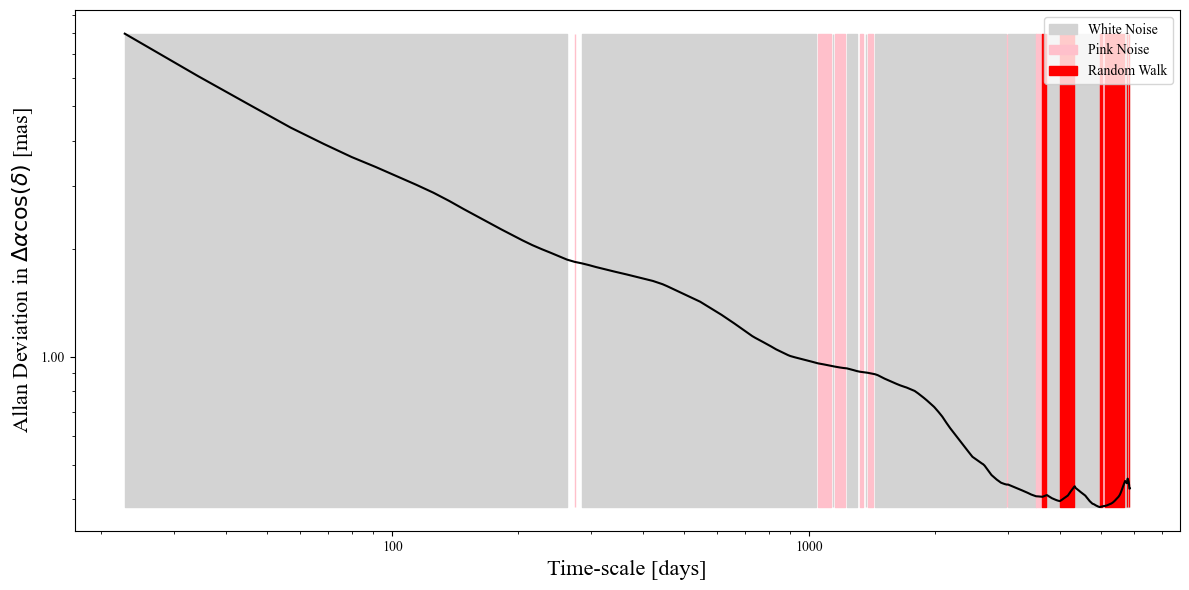}
    \caption{}
         \label{fig:Paris_Allan_ra_cos_dec}
     \end{subfigure}
     \begin{subfigure}[b]{0.49\textwidth}
    \centering
    \includegraphics[width=1\hsize]{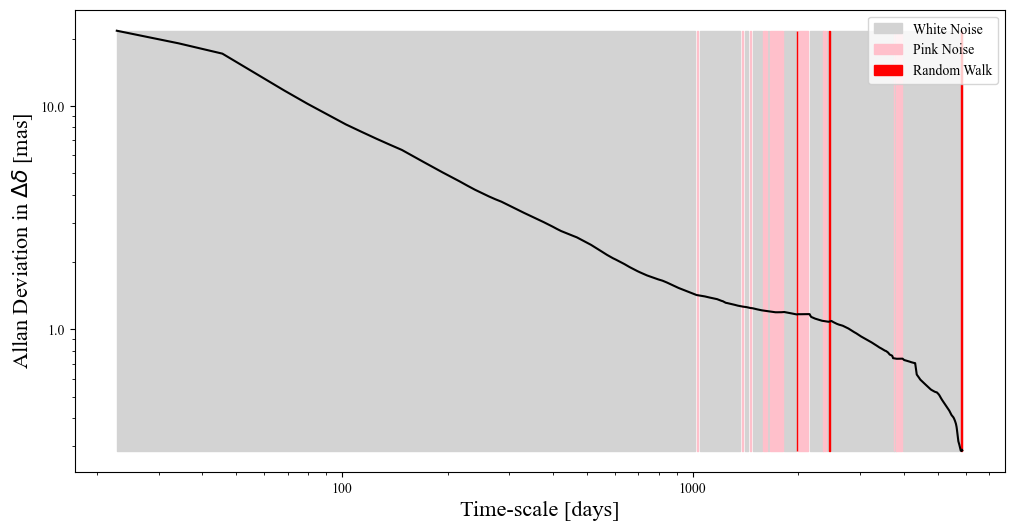}
    \caption{}
         \label{fig:Paris_Allan_dec}
     \end{subfigure}
\caption{\label{fig:Allan} Allan Deviation of quasar 4C31.61. Up: data from Our VLBI analysis. Bottom: data from Paris Observatory Geodetic VLBI Center}
\end{figure}

The Allan deviation of 4C31.61 is shown in Figure \ref{fig:Allan}. They show that white noise dominated, suggesting that the position of quasar 4C31.61 (2201+315) was stable for a short time scale. However, we noted that for a long time scale, there exists random walk, which suggests instability.

Based on Figure \ref{fig:Comparison}, it is evident that there is variability in the quasar's position that can affect the Allan variance results. According to \cite{roland2020}, this variability is suspected to be explained by two types of changes in the quasar nucleus. First, there may be ejections from the quasar nucleus that result in jets.  Second, there is a possibility of a Binary Black Hole (BBH) in the quasar nucleus. Other factors can be considered as well, such as the instrument, atmospheric condition, etc.

\section{Conclusions}
In this study, the position of the quasar 4C31.61 (2201+315) has been obtained by processing VLBI observations, particularly for data spanning from 1990 to 2023.  The quasar's position results obtained using the VieVS software are reasonably similar to those obtained from the Paris Observatory data set, albeit with lower quality. We performed the analysis of the stability of the quasar object 4C31.61 (2201+315) by using Allan variance technique. We found that this quasar exhibits stable position with the dominance of white noise in short time scale. However, in long time scale, the dominance is random walk, which indicate instability.

\section*{References}
\bibliography{iopart-num}

\providecommand{\newblock}{}
\begin{thebibliography}{10}
\expandafter\ifx\csname url\endcsname\relax
  \def\url#1{{\tt #1}}\fi
\expandafter\ifx\csname urlprefix\endcsname\relax\def\urlprefix{URL }\fi
\providecommand{\eprint}[2][]{\url{#2}}

\bibitem{sovers1998astrometry}
Sovers O~J, Fanselow J~L and Jacobs C~S 1998 {\em Reviews of Modern Physics\/}
  {\bf 70} 1393

\bibitem{schuh2012vlbi}
Schuh H and Behrend D 2012 {\em Journal of geodynamics\/} {\bf 61} 68--80

\bibitem{huda2021measuring}
Nurul~Huda I, Hidayat T, Dermawan B, Lambert S, Liu N, Leon S, Fujisawa K,
  Yonekura Y, Sugiyama K, Hirota T {\em et~al.\/} 2021 {\em Experimental
  Astronomy\/} {\bf 52} 141--155

\bibitem{ma98}
Ma C, Arias E, Eubanks T, Fey A, Gontier A~M, Jacobs C, Sovers O, Archinal B
  and Charlot P 1998 {\em The Astronomical Journal\/} {\bf 116} 516

\bibitem{ICRF2}
Fey A~L, Gordon D, Jacobs C~S, Ma C, Gaume R~A, Arias E~F, Bianco G, Boboltz
  D~A, B{\"o}ckmann S, Bolotin S, Charlot P, Collioud A, Engelhardt G, Gipson
  J, Gontier A~M, Heinkelmann R, Kurdubov S, Lambert S, Lytvyn S, MacMillan
  D~S, Malkin Z, Nothnagel A, Ojha R, Skurikhina E, Sokolova J, Souchay J,
  Sovers O~J, Tesmer V, Titov O, Wang G and Zharov V 2015 {\em The Astronomical
  Journal\/} {\bf 150} 58
  \urlprefix\url{https://dx.doi.org/10.1088/0004-6256/150/2/58}

\bibitem{ICRF3}
{Charlot} P, {Jacobs} C~S, {Gordon} D, {Lambert} S, {de Witt} A, {B{\"o}hm} J,
  {Fey} A~L, {Heinkelmann} R, {Skurikhina} E, {Titov} O, {Arias} E~F, {Bolotin}
  S, {Bourda} G, {Ma} C, {Malkin} Z, {Nothnagel} A, {Mayer} D, {MacMillan} D~S,
  {Nilsson} T and {Gaume} R 2020 {\em Astronomy and Astrophysics\/} {\bf 644}
  A159 (\textit{Preprint} \eprint{2010.13625})

\bibitem{gontier2001stability}
Gontier A~M, Le~Bail K, Feissel M and Eubanks T 2001 {\em Astronomy \&
  Astrophysics\/} {\bf 375} 661--669

\bibitem{feissel2003selecting}
Feissel-Vernier M 2003 {\em Astronomy \& Astrophysics\/} {\bf 403} 105--110

\bibitem{gattano2018extragalactic}
Gattano C, Lambert S and Le~Bail K 2018 {\em Astronomy \& Astrophysics\/} {\bf
  618} A80

\bibitem{bohm2018vienna}
B{\"o}hm J, B{\"o}hm S, Boisits J, Girdiuk A, Gruber J, Hellerschmied A,
  Kr{\'a}sn{\'a} H, Landskron D, Madzak M, Mayer D {\em et~al.\/} 2018 {\em
  Publications of the Astronomical Society of the Pacific\/} {\bf 130} 044503

\bibitem{altamimi2023itrf2020}
Altamimi Z, Rebischung P, Collilieux X, M{\'e}tivier L and Chanard K 2023 {\em
  Journal of Geodesy\/} {\bf 97} 47

\bibitem{egbert2002efficient}
Egbert G~D and Erofeeva S~Y 2002 {\em Journal of Atmospheric and Oceanic
  technology\/} {\bf 19} 183--204

\bibitem{wijaya2013atmospheric}
Wijaya D~D, B{\"o}hm J, Karbon M, Kr{\`a}sn{\`a} H and Schuh H 2013 {\em
  Atmospheric effects in space geodesy\/}  137--157

\bibitem{bizouard2019iers}
Bizouard C, Lambert S, Gattano C, Becker O and Richard J~Y 2019 {\em Journal of
  Geodesy\/} {\bf 93} 621--633

\bibitem{allan66}
Allan D~W 1966 {\em Proceedings of the IEEE\/} {\bf 54} 221--230

\bibitem{roland2020}
Roland J, Gattano C, Lambert S and Taris F 2020 {\em Astronomy \&
  Astrophysics\/} {\bf 634} A101

\end{thebibliography}

\end{document}